\documentclass[twocolumn,pra, aps,superscriptaddress]{revtex4-1}
\usepackage{amssymb}
\usepackage{gensymb}
\usepackage{amsmath}
\usepackage{graphicx}
\usepackage{dcolumn}
\usepackage{bm}
\usepackage{epsfig}
\usepackage[usenames]{color}
\usepackage{amsthm} 
\usepackage[colorlinks=true,linkcolor=blue,citecolor=blue,urlcolor=blue]{hyperref}
\usepackage{setspace}
\usepackage{color}
\newcommand{\figref}[1]{Fig.~\ref{#1}}
\renewcommand{\eqref}[1]{Eq.~\ref{#1}}

\let\baraccent=\=

\renewcommand{\v}[1]{\ensuremath{\mathbf{#1}}}

\renewcommand{\=}[1]{\stackrel{#1}{=}}

\newcommand{\matrixel}[3]{\left< #1 \vphantom{#2#3} \right| #2 \left| #3 \vphantom{#1#2} \right>}

\newcommand{\al}[1]{\begin{align}#1\end{align}}
\newcommand{\bs}{\begin{split}}
\newcommand{\es}{\end{split}}

\newcommand{\kp}{$\v k \cdot \v p~$}

\def\be{\begin{equation}}
\def\ee{\end{equation}}
\newcommand{\STO}{SrTiO$_3$~}
\newcommand{\OTS}{SrTiO$_3$}
\newcommand{\LAOSTO}{LaAlO$_3$/SrTiO$_3$~}

\newcommand{\LAO}{LaAlO$_3$~}

\begin{document}
\title{Strain Engineering of the Intrinsic Spin Hall Conductivity in a \STO Quantum Well}
\author{C\"{u}neyt \c{S}ahin}
 \email{cuneytsahin@mailaps.org}
\affiliation{Optical Science and Technology Center and Department of Physics and Astronomy, University of Iowa, Iowa City, Iowa 52242, USA}

\author{Giovanni Vignale}
\affiliation{Department of Physics and Astronomy, University of Missouri, Columbia, Missouri 65211, USA}

\author{Michael E. Flatt\'e}
\email{michael\_flatte@mailaps.org}
\affiliation{Optical Science and Technology Center and Department of Physics and Astronomy, University of Iowa, Iowa City, Iowa 52242, USA}
\date{\today}
\begin{abstract}
The intrinsic spin Hall conductivity of a two-dimensional gas confined to \OTS, such as occurs at an \LAOSTO interface, is calculated from  the Kubo formula. The effect of strain in the [001] (normal to the quantum well direction) and the [111] direction is incorporated into a full tight-binding Hamiltonian. We show that  the spin-charge conversion ratio can be significantly altered through strain and gate voltage by tuning the chemical potential. Strain direction is also a significant factor in the spin Hall response as this direction affects the alignment of the conduction bands.

\end{abstract}
\pacs{pacs numbers}
\maketitle

\section{Introduction}
Two-dimensional electron gases (2DEGs) at  oxide interfaces have attracted enormous interest due to their high carrier density and opportunities for control through atomic-scale interface engineering\cite{Hwang2012}. One of the most prominent examples is the $n$-type conducting interface of the perovskite insulators \LAO (LAO) and \STO (STO) \cite{Ohtomo2004} with high-density and high-mobility electrons. This system supports a rich spectrum of functionalities that can be accurately designed, tuned, and used in applications mainly due to the strongly correlated $d$-orbital electrons of titanium. Observed features include large Rashba coefficients, tunable by field effects and strain\cite{Caviglia2010,Narayanapillai2014}, metal-insulator transitions and multiferroicity \cite{Chakhalian2012,Pena2001},  substantial spin-charge conversion\cite{Jin2016,Lesne2016}, and superconductivity adjustable by an applied gate voltage \cite{Reyren2007,Gariglio2009,BenShalom2010}.  Furthermore, when the Rashba effect is suppressed, very long spin lifetimes have been predicted\cite{Sahin2014} and inferred from room-temperature  spin transport lengths of the order of several hundred nanometers\cite{Ohshima2017} at room temperature.

The spin Hall effect (SHE) describes the emergence of a perpendicular spin current in response to an external electric field in robust spin-orbit coupling systems  \cite{Kato2004b,Wunderlich2005, Engel2007, Murakami2005,Vignale2010}. The spin Hall conductivity (SHC) is the ratio of the spin current to the external electric field. This type of response may originate from different factors, including  extrinsic effects such as skew scattering and side jump, but also may stem intrinsically from the band structure and Berry curvature of the Brillouin zone. Novel materials with giant SHC\cite{Mellnik2014,Sahin2015,Guo2008}, as a result of the spin-orbit interaction, may be very useful in generating and controlling spin currents without external magnetic fields or ferromagnetic contacts. Additionally, strain may significantly influence the band structure affecting the transport properties of the interfacial electron gas. The electron mobilities of \STO can be enhanced up to 300\% under compressive strain\cite{Jalan2011}.  Strain may also alter the critical thickness of the \LAO required to form an electron gas \cite{Bark2011}, at the price of reducing the electric conductivity \cite{Huang2014}. The charge carrier density and the localized magnetic moment at the interface \cite{Nazir2015}, as well as the dielectric response\cite{Antons2005} and the  effective masses \cite{Wunderlich2009} are other strain dependent properties. Therefore,  realistic theories of such materials should consider epitaxial strain as a significant feature of the structure. There have been several attempts to measure spin-charge conversion ratios of these 2DEGs with some impressive results, such as spin Hall angles between 0.15\cite{Jin2016} and 6.3\cite{Wang2017} at room temperature and high spin Hall angles with tunable Rashba coupling \cite{Lesne2016},  which exceeds  the  spin Hall angles of materials such as Pt\cite{Ando2008},  Ta \cite{Morota2011}, and III-V semiconductors \cite{Sih2005}. 
A robust spin-galvanic effect exhibiting a sign change has been predicted \cite{Seibold2017} within a minimal three-band model.  Giant spin-orbit torques,  spin accumulation \cite{Zhang2016}, and Fermi energy-dependent spin responses \cite{Zhou2015} are expected as a result of Rashba  spin-orbit interactions\cite{Hayden2013}.  This large body of work suggests a significant spin-dependent response to electric fields in these systems. However, the intrinsic spin Hall conductivity due to atomic spin-orbit interactions has not been studied in detail, especially considering the  effects of epitaxial and external strains on the intrinsic SHC. 

Here we calculate the intrinsic SHC for a strained two-dimensional electron gas at the \LAOSTO interface from the Kubo formula by a full  Slater-Koster  tight-binding Hamiltonian. This atomistic approach enables a full Brillouin zone calculation of the SHC, thus improving upon perturbative calculations based on   \kp models. Tight-binding Hamiltonians require a small number of parameters and result in far shorter computational times than typical for { \it ab initio} computations. The effective strain along the [001] and [111] directions enters into the Hamiltonian through modified overlap integrals according to bond angles and bond lengths following Harrison's law, which states that overlap integrals  change  by the square of the ratio of unstrained and strained bond length, {\it i.e.} $ H_{hop} \propto (d_{unstr}/d_{str})^2 $. The intrinsic SHC  of these systems is highly sensitive to the chemical potential and also to the strength and direction of the strain, offering opportunities for performance enhancement through strain engineering.

\section{Formalism}

\subsection{Intrinsic Spin Hall Conductivity}
The intrinsic SHC is a result of an interplay between the details of the band structure, the strength of the spin-orbit interaction, the chemical potential and the direction of the current relative to crystal axes\cite{Dyakonov1971,Hirsch1999,Murakami2003,Kato2004b}. For a system with an electric field oriented along  $\hat x$, the spin current is directed along the $\hat y$, and the spin direction is along $\hat  z$. The  spin Hall conductivity can be evaluated from  the Kubo formula as a spin current-electric current response function in the clean static limit \cite{Guo2008}
\al{\label{eq:kubo}
	\sigma_{yx}^z=\frac{e\hbar}{V}\sum_{\v k}\sum_n f_{\v k n}\Omega_n^z (\v k),
 }
where V is the volume of the system, $f_{\v k n}$ is the Fermi-Dirac distribution function, and  the ``Berry curvature" $\Omega_n^z(\v k)$ is
\al{\label{eq:berry}
	\Omega_n^z (\v k) =\sum_{n \neq n'} 2 {\rm Im} \frac{\matrixel{u_{n \v k}}{\hat j_y^z} {u_{n' \v k}} \matrixel{u_{n' \v k}}{\hat v_x}{u_{n \v k}}}{(E_{n\v k} -E_{n' \v k})^2}.
}
The spin current and velocity current operators, $\hat J_x^z$ and $\hat v_y$,  are
\al{ \label{eq:spincurrent}
	J_y^z=\frac{\hbar}{4}(\hat v_y\sigma_z+\sigma_z\hat v_y) ~~\text{and}~~ \hat v_i=\frac{1}{\hbar}\nabla_{k_i}\hat H.
}

Notice that $\Omega_n^z(\v k)$, as defined above, is not a Berry curvature in the strict sense, since the spin current cannot be rigorously expressed as the derivative of the Hamiltonian with respect to a Bloch wave vector.  Instead $\Omega_n^z(\v k)$ is the derivative of the Hamiltonian with respect to a spin-dependent vector potential. Nevertheless, Eq.~(\ref{eq:kubo}) is exact, as it follows from the Kubo formula.  In what follows, we will continue to refer to $\Omega_n^z(\v k)$ loosely as a ``Berry curvature", and we will describe its structure as a function of Bloch wave vector and energy.
It is useful to rewrite Eq.~(\ref{eq:kubo}) so that the chemical potential dependence is captured efficiently by introducing the density of curvatures $ \rho_{\rm doc}(\epsilon)$, which is the contribution of the Berry curvature per unit energy. Introducing the energy-dependent Fermi function $f(\epsilon)$ yields
\al{\label{eq:curvaturedensity}
	\sigma_{yx}^z=\frac{e\hbar}{A} \int d\epsilon \rho_{\rm doc}(\epsilon)f(\epsilon),
}
where A is the area of the two dimensional system. This quantity, the density of Berry curvature, $ \rho_{\rm doc}(\epsilon)$, allows one to interpret the sources of the spin Hall conductivity and its dependence on temperature, external effects such as strain, and the chemical potential. Equations~(\ref{eq:kubo}), (\ref{eq:berry}), and (\ref{eq:curvaturedensity}) suggest that a Hamiltonian which captures wavefunctions, energies and curvatures of the system is required to compute the intrinsic SHC.

\begin{figure}
	\centering
	\includegraphics[width=.5\textwidth]{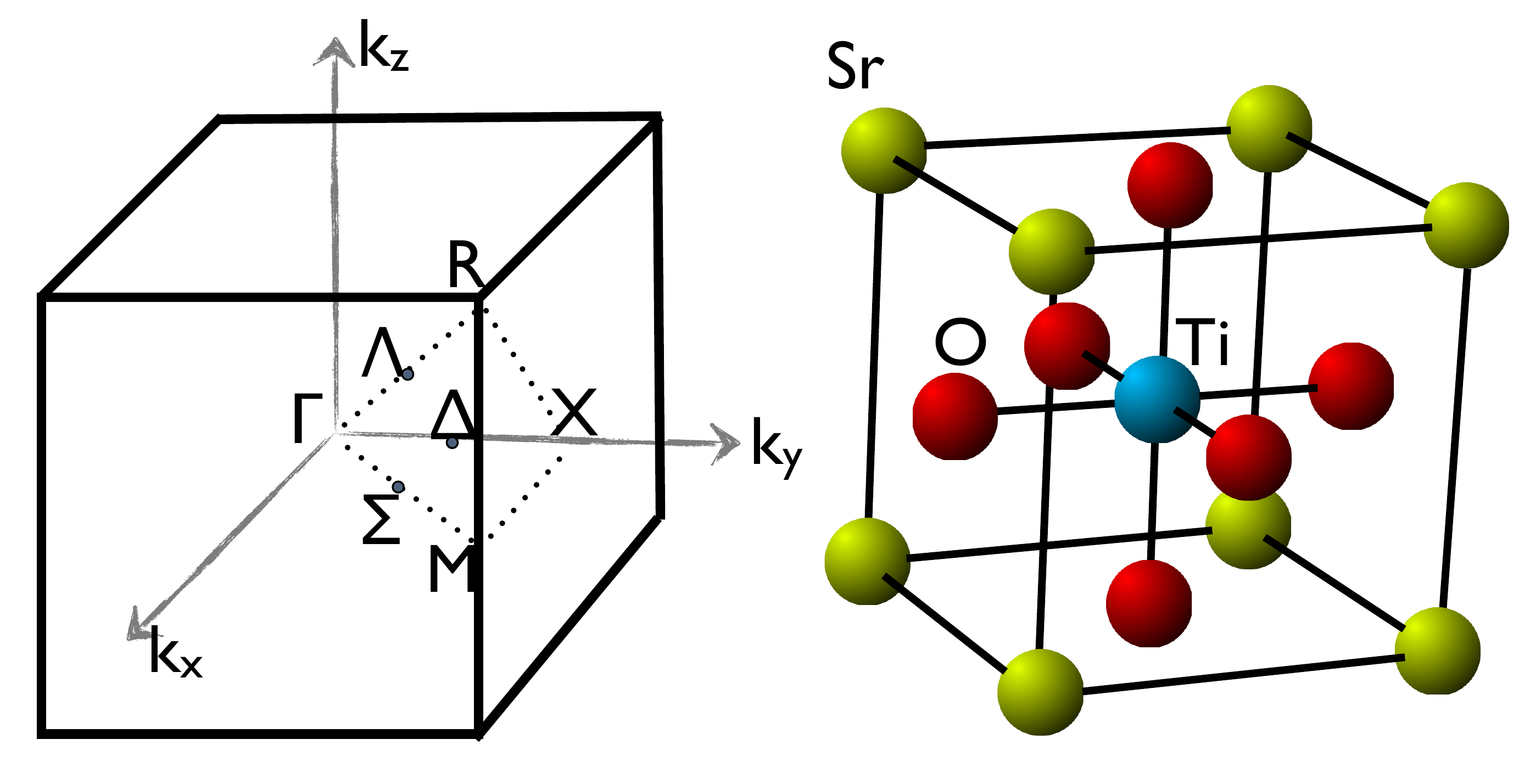} 
	\caption { The reciprocal lattice and the crystal structure of a typical perovskite oxide with the general formula ABO$_3$. Atom B (titanium in this case) is connected to six oxygen atoms forming an octahedron. Atom A at the corners (strontium) usually contributes its \textit{s} electrons that are at high energies. The oxygens' \textit{p} orbitals constitute the valence band. The itinerant $d$-orbitals of Atom B form the conduction band and determine most of the transport properties of $n$-type systems.
	}\label{fig:perovskite}
\end{figure}

\subsection{Strained Tight-Binding Hamiltonian}

Both \LAO and \STO have cubic symmetry and belong to  O$_h^1$ space group\cite{Cardona1965}. A plot of a simple cubic perovskite crystal and its Brillouin zone can be seen in \figref{fig:perovskite}. The existence of both inversion and time-reversal symmetry results in doubly degenerate bands, different from III-V semiconductors and their heterostructures. To get an accurate picture of the wave functions and energies in Eq.~(\ref{eq:berry}) we rely on a low-energy effective Hamiltonian that is constructed using a Slater-Koster tight-binding model\cite{Kahn1964} with first nearest neighbor interactions. 
Starting with an atomic orbital $\phi(\v r -\v R_i)$ at the atomic position $\v R_i$, the Bloch sum of these atomic orbitals is 
\al{
	\psi_n (\v r) =\sum_{\v R_i} e^{i \v k\cdot \v R_{i}}\phi(\v r -\v R_i).
}
The tight-binding Hamiltonian is then calculated by summing over the nearest neighbors at $\v R_j$
\al{\label{eq:TB}
	\hat H_{mn}=\sum_{\v R_i} e^{i \v k\cdot (\v R_j -\v{R_i})} \int \psi_n^*(\v r -\v R_i) H \psi_m(\v r -\v R_j) d{\bf r}
}
The first expression above is the phase factor depending on the relative distances between atoms in the crystals, whereas the integral (also called the overlap integral) depends on  the bond angles and bond lengths. Strain produces two significant changes to the tight-binding Hamiltonian in Eq.~(\ref{eq:TB}). First, it changes atomic distances in the crystal, thus altering the strength of the overlap integrals and phase factors. Second, changes in the bond angles   may induce further two-center Slater-Koster integrals that were not present in the original Hamiltonian due to symmetry. The former is  integrated into the unstrained Hamiltonian via Harrison's scaling law \cite{Froyen1979}, which alters the strength of the interaction in proportion to the bond length and the inverse square rule ($d^{-2}$ rule). The latter is incorporated in the Hamiltonian by changing the directional cosines. For instance, three primitive lattice vectors of perovskite oxides are $\v a_i=\frac{a}{2} \hat i$ and six oxygen atoms around the titanium are located at $ \pm \v a_i $ where $i$ stands for $x$, $y$, or $z$, and $a$ is the lattice constant as seen in Fig.~\ref{fig:perovskite}. For a symmetric general strain $\epsilon_{ij}$,  oxygens in  \figref{fig:perovskite} move to 
\al{
	\v a_1 '&= \frac{a}{2}(\epsilon_{xx}+1, \epsilon_{xy},\epsilon_{xz}), \nonumber\\
	\v a_2 '&= \frac{a}{2}(\epsilon_{yx},\epsilon_{yy}+1,\epsilon_{yz}),\\
	\v a_3 '&= \frac{a}{2}(\epsilon_{zx}, \epsilon_{zy},\epsilon_{zz}+1),\nonumber
}
whereas titanium's position remains unchanged at the center. The distance between titanium and oxygen atoms changes from $d=a/2$ to 
\al{
	d_i'=\frac{a}{2}\sqrt{(1+\epsilon_{ii})^2+\epsilon_{ij}^2+\epsilon_{ik}^2}.
}
In the case of a small strain, this distance reduces to $a/2(1+\epsilon_{ii})$ and the volume of one unit cell changes from $ \Omega_0$ to $ \Omega'=\Omega_0(1+Tr(\epsilon)) $. Without any deformation due to strain, a typical Hamiltonian matrix element between a $d_{xy}$ orbital of titanium and a $ p_{x} $ orbital of the second oxygen is
\al{\label{eq:TBmatrixelement}
	H_{d_{xy},p_x}=2i\sin(\frac{a}{2}  k_y)(pd\pi).
}
where $(pd\pi)$ refers to the overlap matrix element in a $\pi$-bond configuration.
This matrix element transforms under a general strain to
\al{\label{eq:TBmes}
	H_{d_{xy},p_x}=& \nonumber  [ \frac{\sqrt{3}\epsilon_{yx}^2(1+\epsilon_{yy})}{(\epsilon_{yx}^2+(1+\epsilon_{yy})^2+\epsilon_{yz}^2)^{3/2}} pd\sigma ' \\&  \nonumber + \frac{(1+\epsilon_{yy})\ (1-2\epsilon_{yx}^2)}{(\epsilon_{yx}^2+(1+\epsilon_{yy})^2+\epsilon_{yz}^2)^{3/2}}pd\pi ' ]\\
	& \times 2i \text{sin} \left(\frac{a}{2} (k_x \epsilon_{xy}+k_y (\epsilon_{yy}+1)+k_z \epsilon_{yz})\right).
}
where $pd\pi ' = pd\pi/(1+2\epsilon)$ and $pd\sigma ' = pd\sigma/(1+2\epsilon)$ are scaled overlap integrals (for small strain). Eq.~
(\ref{eq:TBmes}) can be further simplified for small strain, as  second order terms can be neglected. As expected this expression approaches Eq.~(\ref{eq:TBmatrixelement}) as the strain approaches to zero. The other off-diagonal elements of the Hamiltonian have been constructed and studied as a function of strain in a similar fashion. 

\subsection{Spin-Orbit Coupling and  Interfacial Quantum Confinement }
We have also added the intrinsic spin-orbit Hamiltonian, obtained by computing atomic spin-orbit couplings from  atomic spectra  using the Land\'e interval rule. The basis of a tight-binding Hamiltonian needs to be doubled once the spin-orbit coupling is introduced. The Hamiltonian with spin takes the form,
\[
H= 
\begin{matrix}\\\mbox{}\end{matrix}
\begin{pmatrix} H_{\text{tb}} & 0 \\ 0 & H_{\text{tb}} \end{pmatrix} +H_{\text{so}},
\]
where $H_{\text{so}}=\lambda_i \v L \cdot \v S$  in the Russell-Saunders coupling scheme. The form of the spin-orbit Hamiltonian for $p$, $d$, and $f$-orbitals has been published \cite{Jones2009}. Here $\v L$ is the linear momentum operator, $\v S$ is the spin operator, and $\lambda_i $ is the strength of the renormalized atomic spin-orbit coupling. This value is related to the atomic spin-orbit couplings, $\xi_{i}$. $\lambda_i$ differs for $p$ and $d$ orbitals, $\lambda_p$ and $\lambda_d$, and vanishes for $s$ orbitals so  $\lambda_s =0$. The atomic spin-orbit coupling depends on the particular configuration of the $p$ or $d$ electrons \cite{Dunn1961}. For a given atomic ground state configuration a standard term symbol has the form of $^{2S+1}X_J$ where $S$ is total spin, $J$ is total angular momentum and $X$ is a letter depending on $L$ such as it is $S$ for $L=0$, $P$ for $L=1$, and $D$ for $L=2$, {\it etc.} \cite{Fisk1968a}. The value of the atomic spin-orbit coupling can be calculated from the Land\'e interval rule, in other words, from the energy difference for the specific term symbol, which are tabulated \cite{Moore1} such as
\al{\label{eq:Lande}
	\xi_i=\frac{E(J)-E(J-1)}{J}, 
}
where the index $i$ represents p or d orbitals. When more than two J exist one will get multiple $ \xi_{i} $ for each splitting. Since resulting energy intervals are very close to each other we considered the average $ \xi $ as the value of spin-orbit coupling. The relation between the spin-orbit coupling $\lambda$ and the atomic spin-orbit coupling $\xi$,  is obtained through total spin $S$, such that $\lambda_i=2S \xi_i$. The splitting of the energy levels in a crystal can be expressed in terms of the splitting of the spectral lines of atoms such as
\al{
	\Delta_0=\frac{E(J)-E(J-1)}{J} \times (2S)\times \frac{2L+1}{2} \times C_N
}
where $C_N$ is a normalization factor that is $1$ for row 2 elements and $1.56$ for row 3 elements, therefore for oxygen and titanium, respectively \cite{Chadi1977}. This factor is required for several reasons. First, both valence and conduction band edges are not formed from pure $p$ or $d$-orbitals and  include higher order atomic orbitals. Second, Wannier functions of atomic orbitals tend to extend more than the typical size of the Wigner-Seitz cell, which causes a volume effect \cite{Chadi1977}. For instance, the ground state of the carbon is $^3P_0$ with $ S=P=1 $ with $3$ energy levels  and term symbols $^3P_0$, $^3P_1$, and $^3P_2$, respectively. The energy difference $E(J)-E(J-1)$ is measured as 16 cm$^{-1}$ \cite{Moore1}, therefore the atomic spin-orbit coupling from Eq.~(\ref{eq:Lande}) gives $\xi_p=2$~meV, and $\lambda_p =2S \times 2$~meV$ =4 $~meV. The resulting splitting of valence band energies in the crystal is then $\Delta_0=4 $~meV$ \times (2L+1)/2=6$~meV which agrees excellently with the experimental splitting. Similarly, we calculate the spin-orbit couplings of oxygen as $\lambda_p=15.2$ meV and titanium as $\lambda_d=20.1$ meV from the atomic spectra. Consequently, this leads to a splitting of the bands by about $30$~meV consistent with the experimental values. 

Finally, for epitaxially grown strontium titanate films an interfacial quantum confinement effect ($H_i$)   has a significant influence on the conduction bands. The total Hamiltonian of our model becomes
\al{
	H_{tot}=H_{tb}^{str}+H_{so}+H_i ,
}
where $ H_{tb}^{str} $, $ H_{so} $, and $ H_i $ are the strained tight-binding, the spin-orbit, and the interfacial quantum confinement terms respectively. The atomic spin-orbit interactions and quantum confinement effects are especially relevant as they alter the band structures  and band degeneracies substantially. Electronic states of \STO in the vicinity of the conduction band minimum (Brillouin zone center) consist of $d$-orbitals of titanium. The crystal potential splits the bands at $\v k=0$ into sixfold ($ \Gamma_{25'} $ irreducible representation) $t_{2g}$ bands ($d_{xy}$, $d_{yz}$, $d_{zx}$) and fourfold $e_g$ bands ($d_{x^2-y^2}$, $ d_{3z^2 -r^2}$ with $ \Gamma_{12'}$ representation) which are located in higher energies. Spin-orbit coupling separates the lower $t_{2g}$ states by about 30 meV, and additional interfacial confinement breaks the fourfold degeneracy by shifting energies of the $d$ orbitals along $\hat z$, such as $d_{yz}$ and $d_{zx}$ compared to $d_{xy}$, thus resulting in five separate doubly-degenerate bands. 
\begin{figure}
	\includegraphics[width=.5\textwidth]{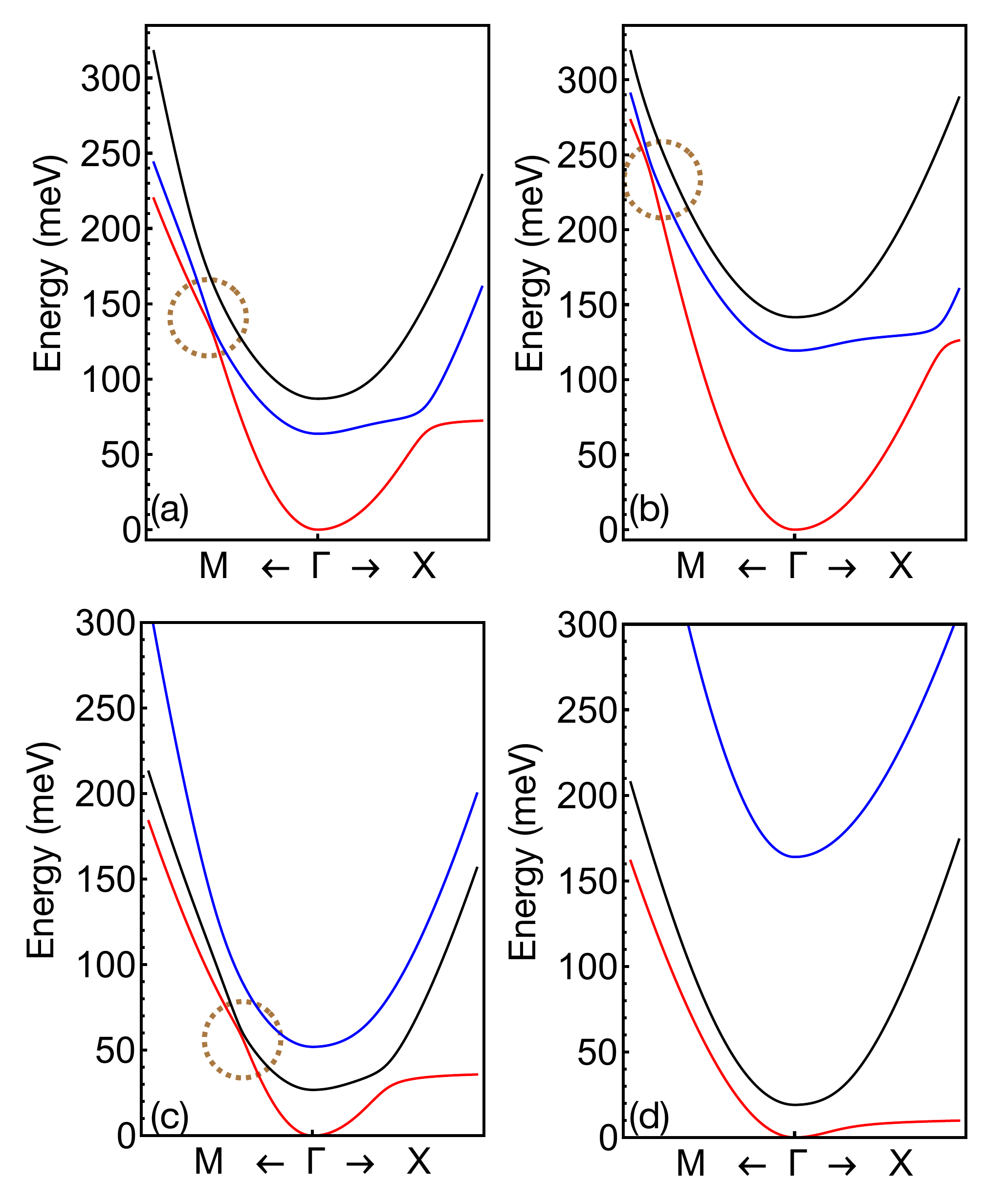} \\
	\caption{(a) The electronic structure of the lowest three conduction bands for a tensile strain of 1.5\% along the [001] direction, (b)  for a compressive strain of -1.5\% along the [001] direction,  (c) for a tensile strain of 0.4\% along the [111] direction, and (d) for a compressive strain of 1.5\% along the [111] direction. The Brillouin zone points where bands nearly touch are circled.}
	\label{fig:bands}
\end{figure}

In \figref{fig:bands} we summarize the two-dimensional band structure of our system for various strains in various directions by plotting the first three conduction bands. In general, the strain shifts certain bands with respect to other ones depending on the strain direction, which we discuss below in Section \ref{sec:results}.

\section{Results and Discussion} \label{sec:results}
\subsection{Stress Along [001] - The Growth Direction}
We first need to address the effect of strain on diagonal matrix elements within the tight-binding Hamiltonian, which correspond to on-site energies. In contrast to the off-diagonal elements of the Hamiltonian, on-site matrix elements have neither directional factors nor overlap integrals (unless further nearest neighbors are added). Strain, however, changes the symmetry of the crystal and as a result can either increase or decrease on-site energies depending on the direction of the strain. For instance, stress along the [001] growth direction induces a biaxial strain and lowers the group symmetry from O$_h$ to its subgroup D$_{4h}$ (nonsymmorphic space group  D$_{4h}^{18}$). This transformation is also analogous to the structural transition of \STO crystals from a cubic to a tetragonal phase at about 100 K \cite{Mattheiss1972a}; consequently, these results would be identical to a low-temperature analysis without strain. The on-site energy shift depends on the magnitude of the strain tensor.
\begin{figure}
	\includegraphics[width=.5\textwidth]{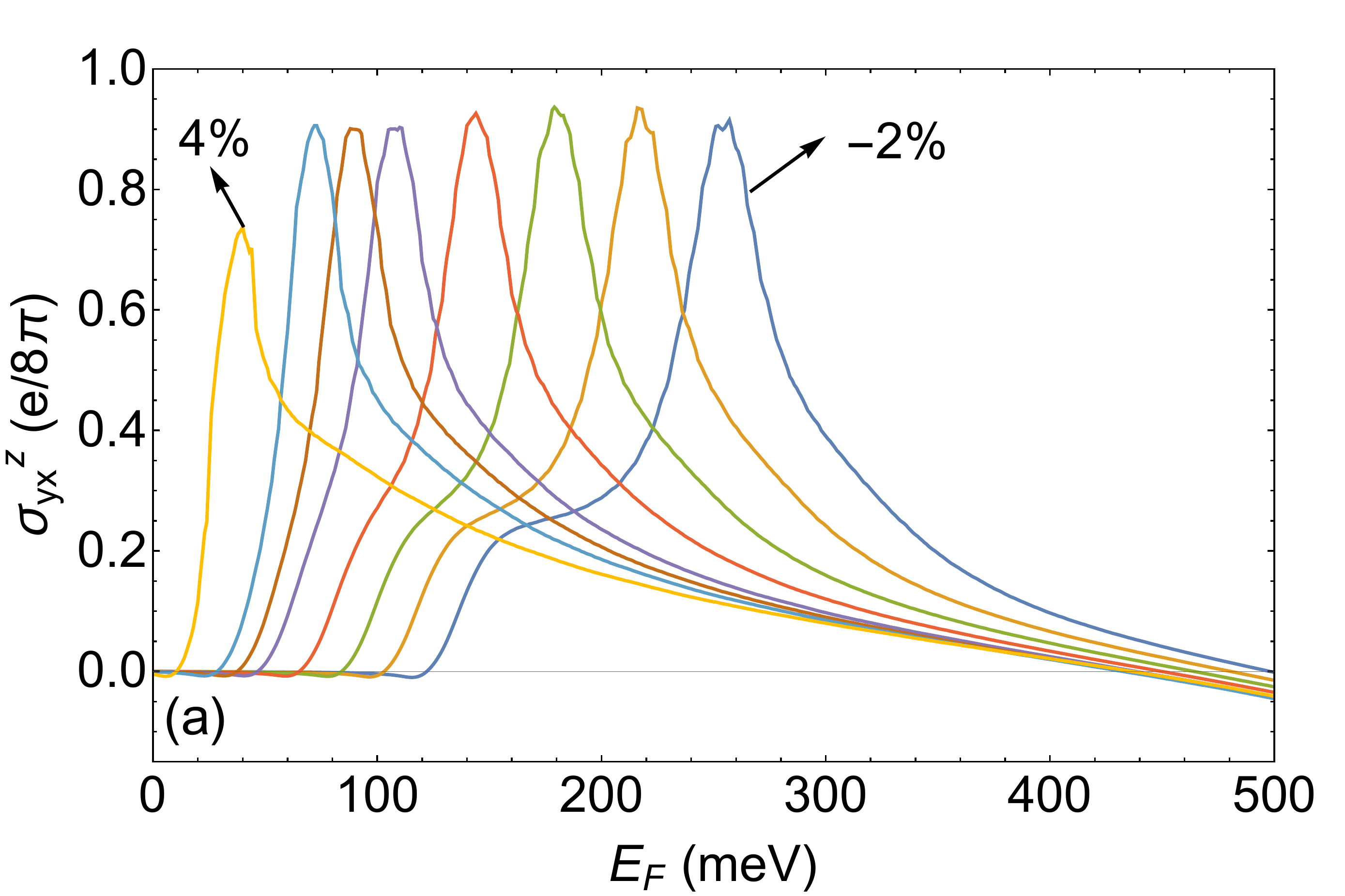} \\
	\includegraphics[width=.5\textwidth]{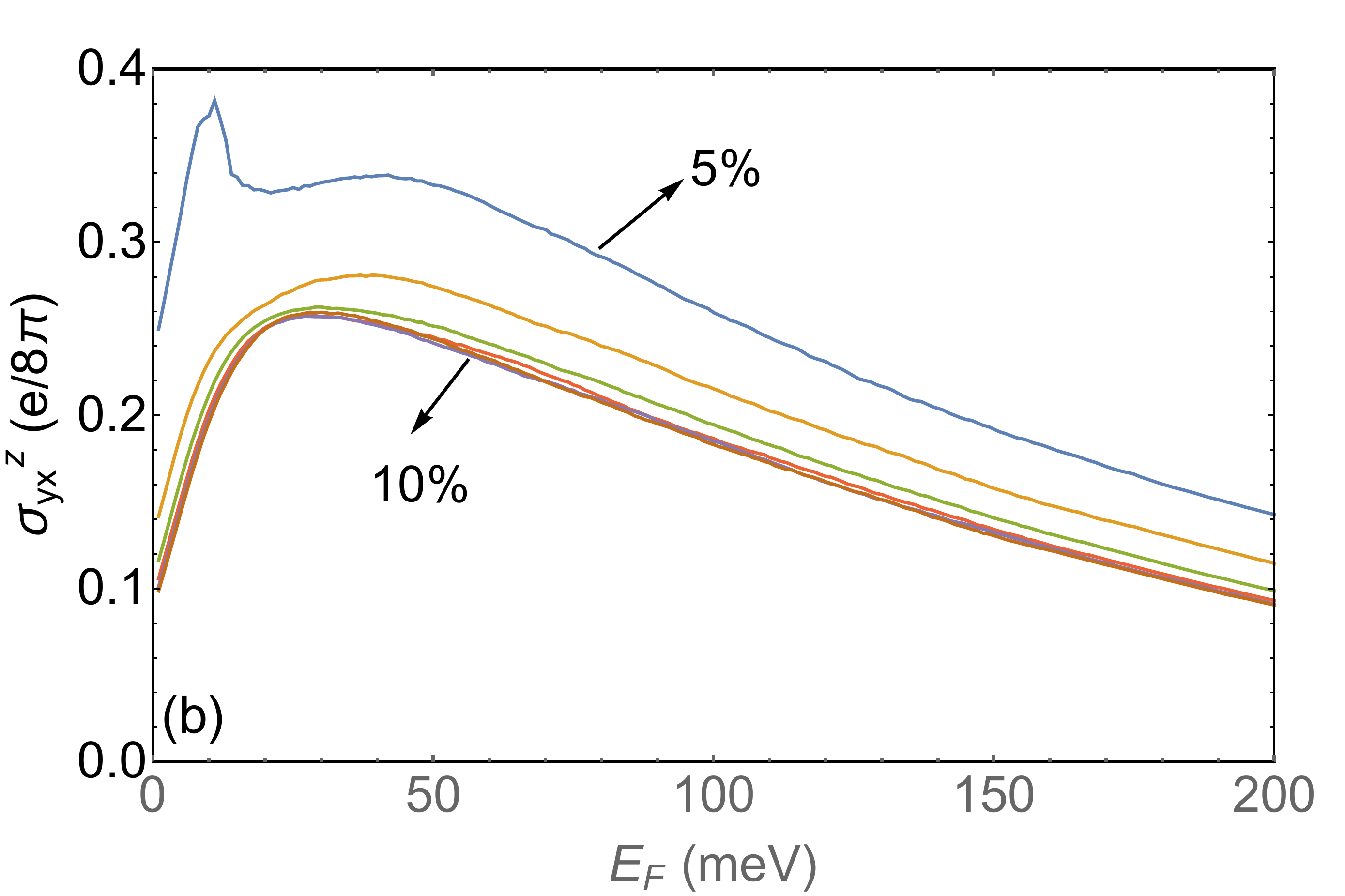} 
	\caption{ Intrinsic SHC as a function of the Fermi level in \LAOSTO 2DEGs for (a) increasing strain on the growth direction [001] from a compressive -2\% to tensile 4\% and (b) for much larger tensile strain from 5\% to 10\%, both by an increment of 1\%.  The zero of the energy corresponds to the conduction band edge at the $\Gamma$ point and the confinement potential is taken to be 100 meV.}
	\label{fig:conf100}
\end{figure}
The relation between strain and stress is determined by the components of the compliance tensor,
\al{
	\epsilon_{ij}=\sum_{k,l} S_{ijkl} \sigma_{kl} 
}
which is a rank 4 tensor but can be greatly simplified for cubic crystals. For a  uniaxial stress along the [001] direction, the stress and strain tensors are related to each other such that
\al{
	\sigma =
	\begin{pmatrix}
		0 & 0 & 0  \\
		0 & 0 & 0  \\
		0  & 0 & 1   
	\end{pmatrix}~~
	\epsilon =
	\begin{pmatrix}
		s_{12}& 0 & 0  \\
		0 & s_{12}& 0  \\
		0  & 0 & s_{11}   
	\end{pmatrix}
}
The relevant elastic constants of the compliance tensor for this study, s$_{11}$, s$_{12}$, and s$_{44}$ (for stress along [111]) are reported in the literature\cite{Poindexter1958}. Our terminology for 1\%  strain is to mean that a stress is applied to generate
 $\epsilon_{zz}=1\%$ along the axis of the stress, and the other strain elements are determined according to force-free boundary conditions on the other surfaces, which follow from the elastic constants, so
$\epsilon_{xx}=\epsilon_{yy}=s_{12}/s_{11} \times1\%$. The conduction bands of \STO at the zone center with $t_{2g}$ symmetry are analogous to valence bands of zinc-blende crystals with a heavy electron, a light electron, and a split-off bands.  Therefore, for a stress along the epitaxial growth direction [001], strain acts as tetragonal crystal distortion, and thus as a perturbation with $\Gamma_{12}$ symmetry. The 3-fold degenerate conduction bands of the strontium with $\Gamma_{25'}$ symmetry will split into doubly degenerate $\Gamma_5^+$ and singly degenerate $\Gamma_4^+$ of the $D_h$ group. Therefore this results in shifting the energy of  $ E_{yz} $ and $ E_{zx}$  with respect to $ E_{xy} $  by $ 3E_{001} $ where
\al{
	E_{001}=2b(\epsilon_{zz}-\epsilon_{xx})
}
and $ b $ is the tetragonal deformation potential. The constant $ b $  (and $ d $ in the case of strain along[111]) is calculated to be -0.51 eV (and -2.15 eV, respectively) \cite{Janotti2011}. Our calculation of the  response of the band edges under tensile and comprensive strains are in excellent agreement with previous {\it ab initio} calculations \cite{Berger2011,Janotti2011}.

Once we introduce the strain along [001] into the tight-binding Hamiltonian by shifting on-site energies and modifying directional cosines and overlap integrals, then we proceed to calculate and plot the intrinsic SHC of the \LAOSTO 2DEG with different configurations in \figref{fig:conf100}. In this configuration, the potential associated with the confinement of electrons is taken as 100 meV. This would be a reasonable estimate since a confinement potential below 30 meV is not adequate to form a 2D electron gas as $d$-electrons would be lost into the bulk \cite{Zhong2013}. Comparison with the electronic band structure, density of curvatures and the SHC calculations leads us to several observations. First, the contribution  from the lowest conduction band is much smaller than that of higher bands until the Fermi level starts to introduce carriers in the  second conduction band. The energy difference between the conduction subbands is large when $E_F$ lies at the conduction band edge, due to the large band gap ($3.2$~eV) and the splitting of these conduction subbands due to strain, the confinement potential, and the spin-orbit coupling. The lowest conduction subband contribution to the SHC is usually negative, leading to a slightly negative SHC up to the Fermi level whereupon the second band starts to contribute,  which suggests  there is a carrier density threshold beyond which the SHC changes sign. This sign change originates from the fact that  negative and positive curvature densities exist at energetically different $k$-points. Although strain doesn't change the energy difference between the second and third bands substantially at the zone center, a compressive strain shifts the first subband away from the higher two subbands, and tensile strain decreases the gap between two. The positive Berry curvature of the second subband therefore contributes at lower Fermi levels for tensile strain, as can be seen from the \figref{fig:conf100}. 

\begin{figure}
	\includegraphics[width=.5\textwidth]{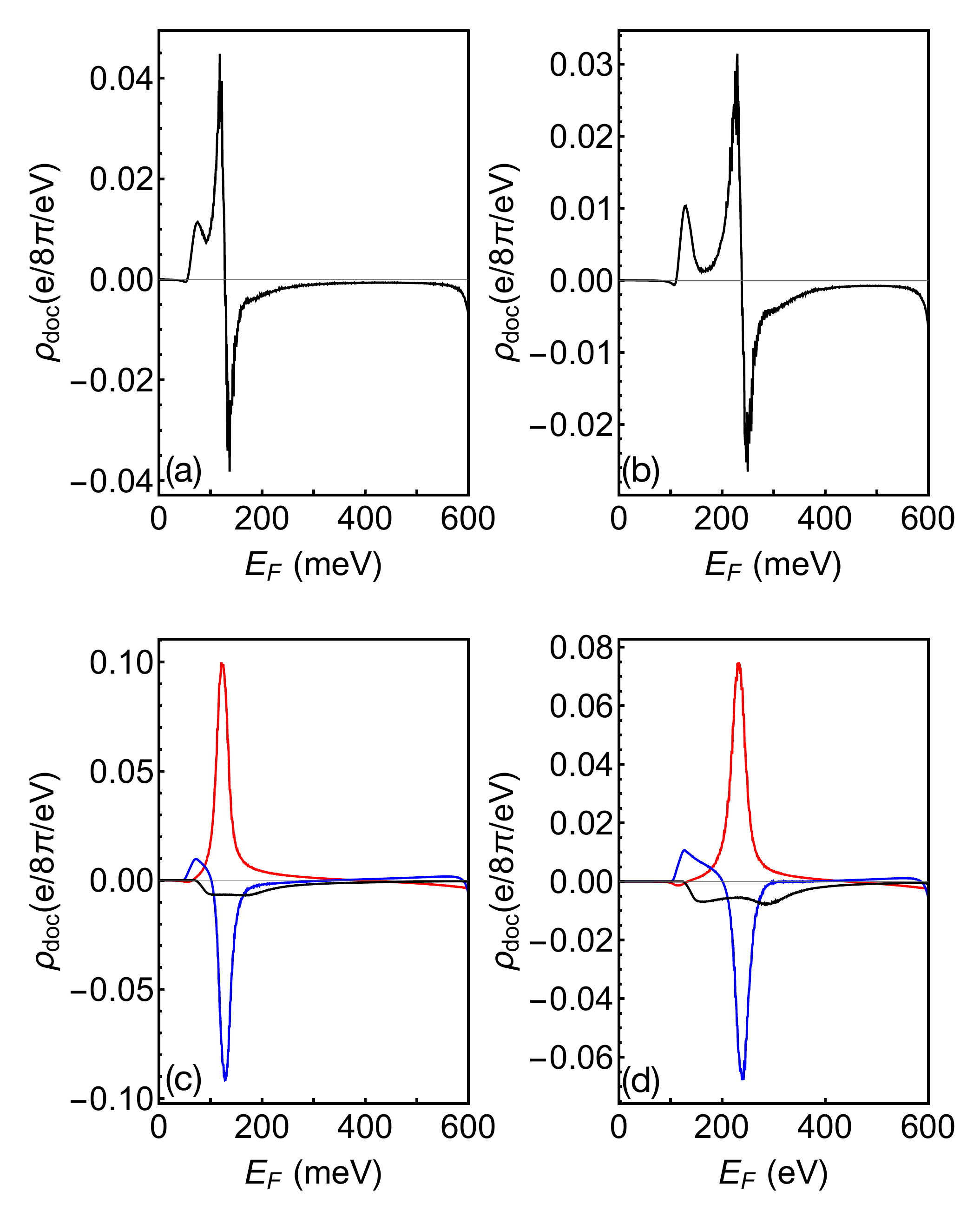}
	\caption { (a) Density of curvatures  for a tensile strain of 1.5\%, (b) density of curvatures for a compressive strain of negative 1.5\%, (c) band resolved contribution to the Berry curvatures for the case of (a), and (d) band resolved curvature contribution for (b) Here red, blue, and black curves represent the first, the second, and the third conduction bands in \figref{fig:bands}, respectively.}
	\label{fig:bandsanddoc}
\end{figure}

The maximum SHC occurs when the Fermi  level crosses the nearly touching first and second bands for strain along [100], as seen in  \figref{fig:bands}(a) and (b). These close band crossings, which are depicted with brown circles, act as sources of very large Berry curvature, and therefore they determine the carrier density at which highest SHC would be observed. Passing through the crossing point, the sign of the Berry curvature is reversed, and, as a consequence, the SHC decreases as Fermi level is further increased.

This result can be understood better by investigating the band structure and distribution of the Berry curvature [Fig.~(\ref{fig:bandsanddoc}(ab)] within the Brillouin zone. Band resolved density of curvature plots provide more insight into this behavior. As shown in Fig.~\ref{fig:bandsanddoc}(cd),  the first conduction band  makes a positive contribution to the SHC, whereas the third band provides negative curvature at all energies. The second conduction band determines the characteristics  the SHC curve. Initially, the second band contributes positive curvatures at small energies. As the chemical potential increases the contribution of the second band  decreases and becomes negative, reaching a magnitude identical to the first band's maximum. The offset in the energies where these maximum and minimum curvatures are located results in the features seen in the total SHC. Comparing different strains [such as 1.5\% in \figref{fig:bandsanddoc}(ac) vs. -1/5\% in (bd)] leads to the conclusion that shifting the bands with strain shifts the chemical potential where the maximal SHC occurs. 

We also calculated the SHC for carrier densities that vary from $1.5 \times 10^{14}$cm $^{-2}$ to $6 \times 10^{14}$ cm $^{-2}$, corresponding to moving  the Fermi level from 90 meV to 300 meV. These densities, which are achievable through doping or gate voltage, are in excellent agreement with previous experiments on strained \LAOSTO 2DEGs \cite{Nazir2015}. Our calculations are also in agreement with the observation that uniaxial tensile strain greatly enhances the carrier density \cite{Nazir2014}. Another effect of strain in this direction is the tetragonal deformation of the octahedral structure consisting of 6 oxygens. This deformation leads to a rotation angle. The distance between titanium and the oxygens in the $xy$-plane $d_\parallel=a_\parallel/\cos(\pi -\alpha)$, where $ \pi/2 -\alpha $ is the angle along Ti-O-Ti. This angle is exactly $\pi/2$ without strain, corresponding  to a completely straight line along the Ti-O-Ti direction. However, this angle reduces with tensile strain, whereas the out of plane distance between titanium and oxygen  ($d_\perp$) increases and the in-plane distance  ($d_\parallel$) decreases. This results in a rotation which can be expressed in terms of strain elements,  as 
$\alpha =\cos^{-1}  [1/(s_{11}/s_{12}\times\epsilon_{zz}+1)]$
For a strain of 1.4\% this effect results in a rotation of 4.6 degrees which is in an excellent agreement with the experimentally measured value of 4.58 degrees in a 2DEG with 300 unit cell  \STO thickness \cite{Huang2014}.

\subsection{Strain along [111]}
Applying strain along [111] differs from [001] as the strain affects a different diagonal element of the Hamiltonian. A uniaxial stress along [111], where $\sigma_{ij}=1$, results in a strain tensor
\al{
	\epsilon =\frac{1}{3}
	\begin{pmatrix}
		s_{11}+2s_{12} & s_{44}/2 & s_{44}/2  \\
		s_{44}/2 & s_{11}+2s_{12} & s_{44}/2  \\
		s_{44}/2  & s_{44}/2 & s_{11}+2s_{12}   
	\end{pmatrix} .
}
\begin{figure}
	\includegraphics[width=.5\textwidth]{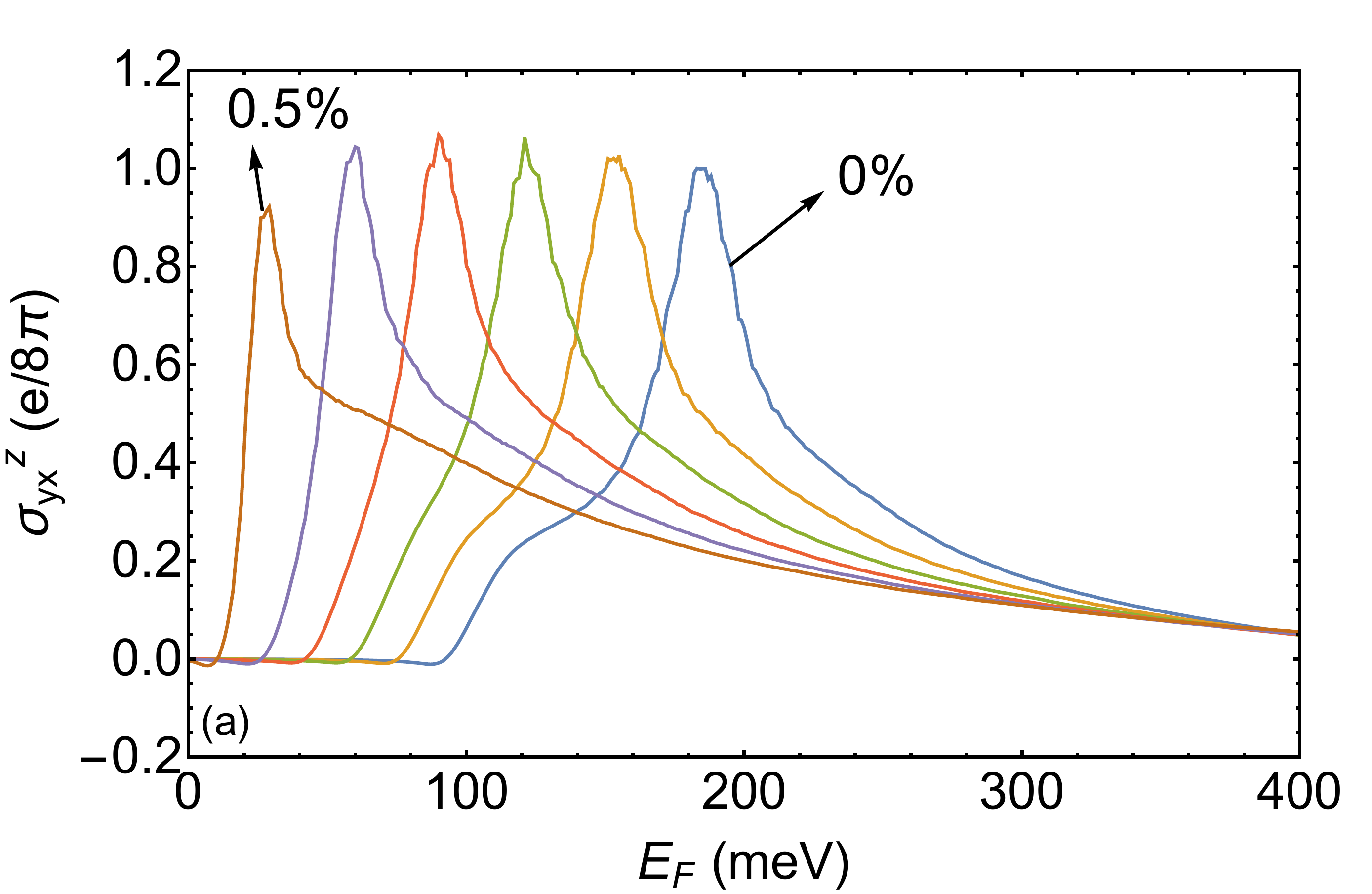} \\
	\includegraphics[width=.5\textwidth]{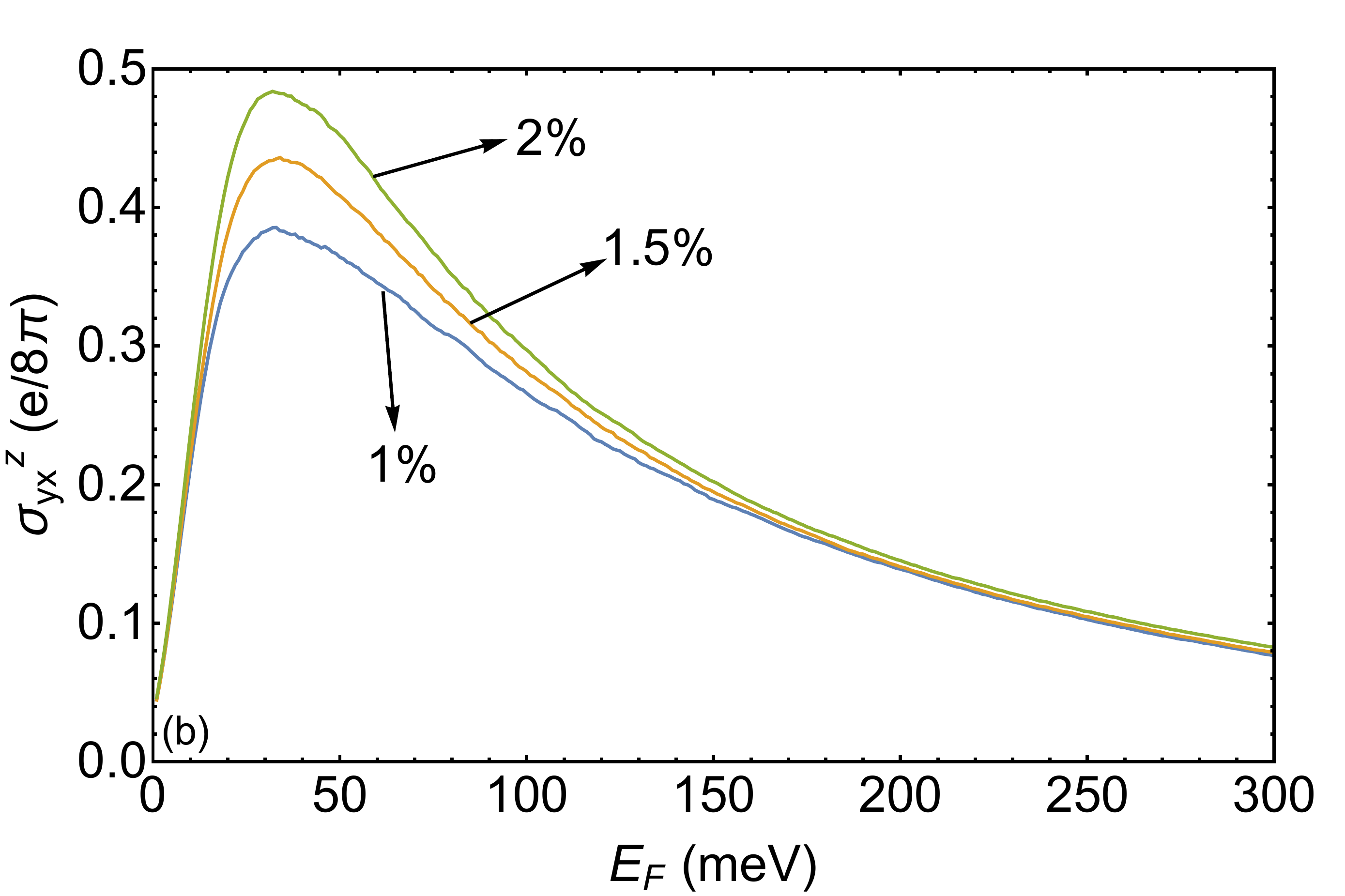} 
	\caption{Intrinsic SHC as a function of the Fermi level in \LAOSTO 2DEGs strained along the [111] direction. (a) tensile strain from 0\%\ to 0.5 \%, increasing  by an increment of 0.1\%, (b)  tensile strain of 1\%, 1.5\%, and 2\%. The energy is measured from the conduction band edge at the $\Gamma$ point. The confinement potential is taken as 100 meV.}
	\label{fig:conf111}
\end{figure}
Here 1\% strain indicates $\epsilon_{xx}=\epsilon_{yy}=\epsilon_{zz}=1\%$. Other elements of the tensor are calculated via compliance tensor elements. This type of strain acts as a perturbation with $\Gamma_{15}$ symmetry which shifts  $ E_{xy} $ by $E_{111} $, where 
\al{
	E_{111}=2\sqrt 3 d \epsilon_{xy},
}
and $d$ is the trigonal (or rhombehedral)  deformation potential. Our calculations of strained band structures are in an excellent agreement with  previous {\it ab initio} computations \cite{Janotti2011}.

A negative strain pushes bands away from each other and the SHC is nearly zero until the doping is increased to the point where the chemical potential crosses to the second conduction band. This high doping would be a difficult doping level to achieve. At the zone center, however, a positive [111]  strain moves the $d_{xy}$ band closer to the upper energy levels (different from  [001] quantum wells where the $d_{xy}$ subband is separated from the higher bands). This would  close the gap from the interfacial potential between the first subband and higher subbands. In the case of positive strain, we observe two distinct behaviors. For  very small strain from 0\% to 0.5\% our results resemble those for strain along the [001] direction, {\it i.e.} increasing strain causes bands to move closer and the chemical potential of  the maximum SHC is also shifted towards the band edge. 

\begin{figure}
	\includegraphics[width=.49\textwidth]{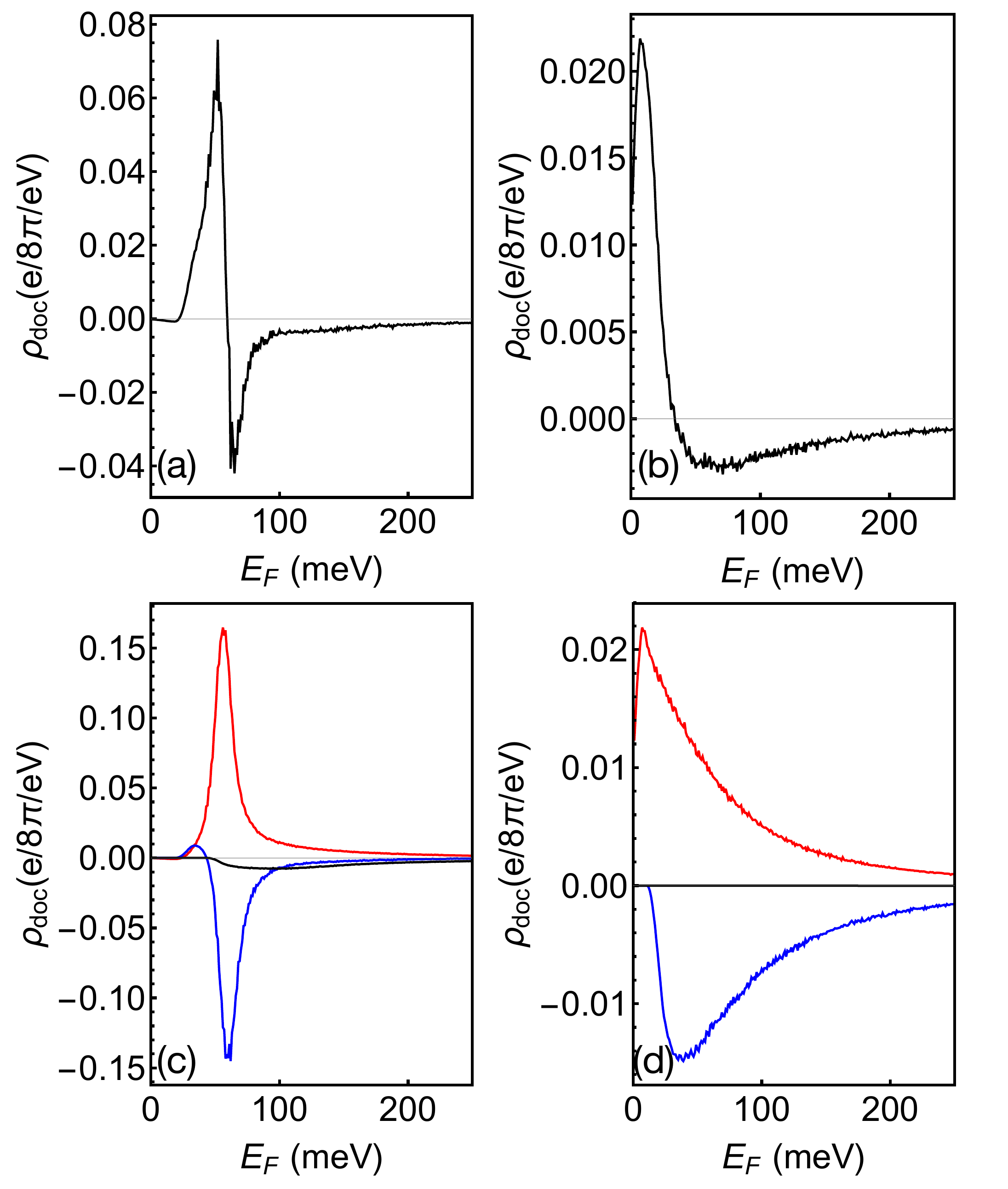}
	\caption{(a) Density of curvatures for a tensile strain of 0.4\% along the [111] direction, (b) for compressive strain of 1.5\%, (c) band resolved contribution to the Berry curvatures for the case of (a), and (d) band resolved curvature contribution for (b). Red, blue, and black curves represent the first, the second, and the third conduction bands.}
	\label{fig:bandsanddoc111}
\end{figure}

The resulting spin Hall conductivity can be seen in \figref{fig:conf111}. However, once the lowest conduction band is increased to the level of the second conduction band and forms a degenerate state at about 0.5\% strain, the overall shape of the SHC changes. As increasing strain increases the separation between bands, we observe a similar behavior as in \figref{fig:conf100}(b). One significant  difference is that increasing strain increases the spin Hall conductivity. This can be explained readily by computing the curvature of the third band, which is negative. Rising strain results in a larger separation between the first two bands and the third band; thus, the negative curvature of that band has less impact on the overall spin Hall conductivity (Fig.~\ref{fig:bandsanddoc111}). Band-resolved density of curvature plots [Fig.~\ref{fig:bandsanddoc111}(cd)] indicate that for values of the strain up to 0.5\% the evolution is similar to what was found for strain in the [001] direction. The first band has primarily positive curvature, the second band has mostly negative curvature but shifted in energy slightly with respect to the first band, which results in a threshold chemical potential at which total curvatures change sign, thus creating a maximum SHC. This threshold Fermi level is about 70 meV above the band edge and corresponds to a doping level with a carrier density of 1.7$\times 10^{15}$cm$^{-2}$ In the case of 1.5\% strain, the third band is shifted so far away that it does not contribute to SHC for achievable chemical potentials. The first band and second band contributions compete with each other as the positive first band has a slightly lower energy than the second band. In this case, the threshold Fermi level is in between the first two conduction bands ($ \approx 40$~meV) and corresponds to a carrier density of 1.5$\times10^{14}$cm$^{-2}$. These carrier densities are in the experimental range. 

\section{Conclusions}
We have developed a  tight-binding Hamiltonian description of \LAOSTO 2DEGs that accounts for strain via changing bond lengths and angles. Spin-orbit coupling and  interfacial quantum confinement is included in the Hamiltonian. We  calculated the intrinsic spin-Hall conductivities as a function of the strain and chemical potential. Our results reveal a strong effect of the strain on the spin Hall conductivities as the doping level changes. We have also investigated the source of the large SHC by plotting the band-resolved density of Berry curvatures, and identified ``hot points''  with exceptionally large Berry curvatures in the Brillouin zone.  Strains along different directions mainly alter the intrinsic SHC through changes in the band structure and the band curvatures. Our calculations also show that the intrinsic SHC of strained systems is of the order of (e/8$\pi$), so the effect is comparable to the values that were calculated from Rashba  spin-orbit interaction \cite{Zhou2015, Hayden2013}.  Exceptionally large, tunable spin Hall conductivities in these two-dimensional systems with high carrier densities and large mobilities suggest that they could play a substantial role in developing spintronic devices.

\section{Acknowledgement}
We acknowledge support of the Center for Emergent Materials, an NSF MRSEC under Award No. DMR-1420451, and an ARO MURI.
\bibliography{central-bibliography}
\end{document}